\documentclass[12pt]{article}

\usepackage{color,graphics,epsfig,graphpap}

\newcommand{\be}{\begin{eqnarray}}
 \newcommand{\ee}{\end{eqnarray}}
 \newcommand{\nee}{\nonumber\end{eqnarray}}
 \newcommand{\nn}{\nonumber\\}
 
  \newcommand{\bc}{\begin{center}}
 \newcommand{\ec}{\end{center}}

\def\m             {\mu}
\def\n              {\nu}

\def\s              {\sigma}
\def\g              {\gamma}

\def\t          {\tau}

\def\red{\color{red}}

\def\blue{\color{blue}}

\usepackage{amsmath,amssymb,amsfonts,amsbsy}
\usepackage{graphicx}
\usepackage{wrapfig}

\begin{document}
\addcontentsline{toc}{subsection}{{The axial form factor and polarization of \\
      the final nucleon in quasi-elastic $\nu-N$ scattering}\\
{\it B.B. Author-Speaker}}

\setcounter{section}{0}
\setcounter{subsection}{0}
\setcounter{equation}{0}
\setcounter{figure}{0}
\setcounter{footnote}{0}
\setcounter{table}{0}

\begin{center}
\textbf{The axial form factor and polarization of \\
      the final nucleon in quasi-elastic $\nu-N$ scattering }
\footnote{The paper is supported by  a priority Grant between Bulgaria and JINR.
S.M. Bilenky acknowledges the support of  AvH Stiftung  (contract Nr. 3.3-3-RUS/1002388)
and RFBR Grant N 13-02-01442; E.Ch.'s work is supported by "HiggsTools" Initial Training
Network of 7-th framework programme of the EC.  }

\vspace{5mm}

S.M.Bilenky$^{\,1\,\dag}$ and \underline{E.  Christova}$^{\,2\,\dag\dag}$,

\vspace{5mm}

\begin{small}
    (1) \emph{JINR, Dubna, Russia } \\
  (2) \emph{Institute for Nuclear Research and Nuclear Energy, Bulgarian Academy of Sciences, \\Sofia, Bulgaria} \\
   \emph{E-mail:} $\dag$\emph{bilenky@gmail.com}; $\dag\dag$\emph{echristo@inrne.bas.bg};
\end{small}
\end{center}

\vspace{0.0mm} 

\begin{abstract}
 We have calculated the polarization of the final
nucleon in charged current quasi-elastic  $\nu -N$
scattering. We show that  the longitudinal and transverse
polarizations, as well as their ratio  exhibit simple dependence on
the axial form factor and their sensitivity to the axial mass  is
much stronger than that of the cross section.
 This  suggests that measurements of the
polarization of the nucleon in the high-statistics 
neutrino experiments  could provide important information on the
axial form factor.
\end{abstract}


\section{Introduction}

Important for understanding the  electromagnetic structure of the nucleon are
 the two, Dirac and Pauli, electromagnetic form factors (FFs) $F_1(Q^2)$ and $F_2(Q^2)$, that determine elastic electron-nucleon
scattering.

There are two ways of extracting $F_{1,2}$, or the more convenient
experimentally charge and magnetic FFs $G_E=F_1-\tau F_2$ and
$G_M=F_1+F_2$, $\t = Q^2/4M^2$.  The standard,  Rosenbluth, procedure
 is based on the unpolarized cross section and  determines $G_E$ and $G_M$ separately
 with a limited sensitivity to $G_E^p$ at higher $Q^2$. It was found:
\vspace{-.2cm}

\begin{enumerate}
 \item  At relatively low $Q^2\lesssim 5\,GeV^2$ the proton and neutron magnetic FFs  exhibit approximately the {\it same} dipole $Q^2$-dependence,
 $G_D(Q^2)=\left(1+Q^{2}/0.71~\mathrm{GeV}^{2}\right)^{-2}$:
 \be
G^{p}_{M}=\mu_{p}~ G_D(Q^2), \quad
G^{n}_{M}=\mu_{n}~ G_D(Q^2).
\ee
\item  Up to $Q^2\lesssim 6~\mathrm{GeV}^{2}$ all
 data exhibit  "{\it scaling}"  of the electric $G_E^p$ and magnetic $G_M^p$ FFs of the proton:
\begin{equation}\label{scaling}
{\cal R}(Q^{2})=\frac{\mu_{p}G^{p}_{E}(Q^{2})}{G^{p}_{M}(Q^{2})}\simeq 1.
\end{equation}
\end{enumerate}
\vspace{-.2cm}

 In the late 90-ies Jefferson Lab. started series of new
type of experiments that allowed a direct  measurement of ${\cal
R}$. In 1968 it was shown \cite{AkhRek} that  the ratio of the
transverse $P_\perp$ and longitudinal $P_\parallel$ polarization of
the recoil proton is directly proportional to $\cal R$:
\begin{equation}\label{ratio}
\frac{P_\perp}{P_\parallel}=-\,\frac{G^{p}_{E}}
{G^{p}_{M}}\sqrt{\frac{2\varepsilon}{\tau(1+\varepsilon)}},
\end{equation}
where $\varepsilon=[1+2(1+\tau)tan^{2}\theta/2]^{-1}$,  $\theta$ is
the scattering angle. JLab measured the ratio $P_\perp /P_\parallel$
 in the energy range $Q^2$=[0.5 -- 8.5] $GeV^2$ and
unexpected results were obtained \cite{Perdrisat}:
\begin{enumerate}

\item  "Scaling" {\it does not} hold.
The form factor  $G_E^p(Q^2)$ decreases much faster than  $G_M^p(Q^2)$: $\cal R$ = 1 at $Q^2 \simeq 1 ~\mathrm{GeV}^{2}$ and falls down to
${\cal  R}= 0.2$ at $Q^{2}=5.6 ~\mathrm{GeV}^{2}$.

\item  There is a clear discrepancy between the two methods in extracting  ${\cal R}$.
\end{enumerate}

Polarization experiments drastically changed our knowledge about the
e.m. FFs and raised the important questions about radiative
corrections and 2-photon exchange.

 The JLab  results  strongly motivated our studies of the recoil
 nucleon polarization in  charged current quasi-elastic (CCQE) $\n (\bar\n )-N$ scattering
 as a source of  independent information about the axial form factor.
 We obtain analytic expressions for the polarization
  and estimate  numerically the  sensitivity  of the polarization and the cross sections
 to the axial mass. Most of the presented results  can be found in more details in~\cite{EChSM}.

 \section{The weak charged current  form factors}

We study the CCQE processes:
\be
\n +p\to \m^+ + n,\qquad \bar\n +n\to \m^- + p\label{process}
\ee
which are the dominant processes at low neutrino energies and give a direct information on the charged current (CC) weak form factors.

The matrix elements of (\ref{process}) is determined by the 4 weak CC FFs -- $F_{1,2}^{CC}$, $G_A$ and $G_P$:
 \be
 {\cal M} &= &\frac{G_F}{\sqrt 2}\, (\bar u_\m \g^m (1\pm \g_5)u_\n )\cdot \langle N'\vert
J_\m^{CC} \vert N \rangle\\
 \langle N'\vert  J_\m^{CC} \vert N \rangle &=& \bar u_{N'}\,\left( \g_\m\,
F_1^{CC}+\frac{i\sigma_{\m\n}q^\n}{2M}\, F_2^{CC}
+\g_\m\,\g_5\,G_A+\frac{q_\m}{2M}\,\g_5\,G_P \right)u_N\label{M}
\ee
Due to CVC  $F_{1,2}^{CC}$ are related to the e.m. FFs:
 \begin{equation}\label{CVC}
F^{CC}_{1,2}(Q^{2})=F^{p}_{1,2}(Q^{2})-F^{n}_{1,2}(Q^{2}),
\end{equation}
where $F^{p}_{1,2}$ and $F^{n}_{1,2}$ are the
Dirac and Pauli  form factors of the proton and  neutron,
 known at present in a wide region of $Q^{2}$ \cite{Perdrisat}.
 The hypothesis for partial conservation of the axial current (PCAC)
 implies that the contribution of $G_{P}(Q^{2})$ can be neglected.  Thus, study of  the CCQE processes
  (\ref{process}) will give information about the axial form factor $G_{A}(Q^{2})$.

  In analogy with the electromagnetic FFs,  $G_A$ is usually parameterized by the dipole formula:
\begin{equation}\label{dipoleGA}
G_{A}(Q^{2})=\frac{g_{A}}{(1+\frac{Q^{2}}{M^{2}_{A}})^{2}}.
\end{equation}
Here  $g_{A}=1.2701\pm 0.0025$   is the axial constant,
known from the neutron $\beta$-decay data
 and $M_{A}$ is a parameter --  the "axial mass". At present,  experiments on measurements of the
 CCQE cross section, performed at different neutrino energies and
 on different nuclear targets suggest different values for $M_A$ \cite{Ransome}:
 \be
d\,\, {\rm or}\,\, H-target\qquad M_{A}&=& 1.03\pm 0.02~\mathrm{GeV}\nn
 Fe-target\qquad M_{A}&=& 1.26 _{-0.10}^{+0.12} {}_{-0.12}^{+0.08}~\mathrm{GeV}, \quad MINOS\nn
 H_20-target\qquad  M_{A}&=& 1.20\pm 0.12~\mathrm{GeV},\quad K2K\nn
 C-target\qquad  M_{A}&=& 1.05\pm 0.02\pm 0.06~\mathrm{GeV},\quad NOMAD\nn
 C-target\qquad  M_{A}&=& 1.35\pm 0.17~\mathrm{GeV},\quad MiniBooNE
 \ee
Though compatible within 2 $\sigma$ errors, these results  show
a clear discrepancy for the central values of $M_A$, that could originate in different reasons.
The precise determination of the axial FF
is important not only for understanding
the nucleon structure, but it is a basic ingredient in  interpretation of the neutrino oscillation experiments.
Here we suggest that measurement of the final nucleon polarization could provide an important independent information about $G_A$.

\section{Polarization of the final nucleon}

 T-invariance implies that the  polarization vector of the final nucleons in (\ref{process}) lays in the scattering
  plane. 
  We define its longitudinal $s_\|$ and transverse $s_\bot$ components:
 \begin{equation}\label{polarLab}
\vec{s}=s_{\bot}\vec{e}_{\bot}+s_{\|}\vec{e}_{\|},
\end{equation}
where $\vec{e}_{\bot}$ and $\vec{e}_{\|}$ are two orthogonal unit vectors in the scattering plane,
 $\vec{e}_{\|}=\vec{p'}/|\vec{p'}|$, $p'$ is the 4-momentum of the final nucleon.
We obtain:

$\bullet$ The transverse polarization exhibits a simple linear dependence on $G_A$:
\be
(J_0\,s_{\bot})^{\n ,\,\bar\n} =
\frac{-\,2E'\,\sin\theta}{\vert\vec q\vert}\left[\pm
y\,G_M^{CC}+(2-y)G_A\right]\,G_E^{CC}\label{trans2}
 \ee

$\bullet$ The longitudinal polarization $s_\|$ is expressed solely in terms of $G_A$
 and $G_M^{CC}$, i.e. the best known magnetic form factors
 of the proton and neutron, the poorly known $G_E^{CC}$ does not enter:
 \be
\left(J_0\,s_{\|}\right)^{\n ,\,\bar\n}=-\, \frac{q_0}{\vert \vec
q\vert}\,\left[ \pm y\,G_M^{CC}+(2-y)\,G_A\right]
 \left[(2-y)\,G_M^{CC}\pm \left(y+ \frac{2M}{E}\right)\,G_A\right].\label{long2}
\ee

$\bullet$ If the neutrino detector is in a magnetic field, then  both $s_{\bot}$ and $s_\|$
  could be measured (like in  elastic $e-p$ scattering).
  Their ratio exhibits a simple linear dependence on $G_A$:
\be
\left(\frac{s_\|}{s_\perp}\right)^{\n
,\bar\n}&=&\frac{q_0}{2E'\sin\theta}\,\, \frac{
\left[(2-y)\,G_M^{CC}\pm \,G_A(y + 2M/E)\right]}{G_E^{CC}}.
\ee

$\bullet$ The quantity  $J_0^{\n ,\,\bar\n}$ is determined via the differential cross section:
\be
J_0^{\n ,\bar\n}=\frac{d\s^{\n ,\,\bar\n}}{dQ^2}\cdot\,\frac{4\pi}{G_F^2},
\ee
and is given by the expression:
 \be
J_0^{\n ,\,\bar\n}
 &=&2(1-y)\left[G_A^2+\frac{\tau (G_M^{CC})^2+(G_E^{CC})^2}{1+\tau}\right]
 +\frac{My}{E}\,\left[\,G_A^2-\frac{\tau (G_M^{CC})^2+(G_E^{CC})^2}{1+\tau}\right]\nn
 &&+y^2\,(G_M^{CC}\mp G_A)^2\pm 4y\,G_M^{CC}\,G_A.
 \ee
Here $y,\, q_0,\,\vert \vec q\vert$ are kinematic factors, $E'$ is
the energy of the final lepton.

\section{Numerical results}

Using the commonly used parametrizations for the e.m. FFs, we examined the sensitivity of
 $s_\|$ and $s_\perp$,
 and their ratio $s_\|/s_\perp$ on the axial mass for the following values of $M_A$:
\be
&&1) \,\,M_A=1.016\, - \,full \,\,(black)\,\,line\nn
&&2)  \,\,M_A=1.20\,  -\, dashed\,\,({\red red})\,\, line\label{MA}\\
&&3)  \,\,M_A=1.35\, -\, dash-dotted \,\,({\blue blue})\,\,line
\nee
We compared it to the sensitivity of the cross section.

\begin{figure}[b!]
  \centering
  \begin{tabular}{ccc}
    \includegraphics[width=50mm]{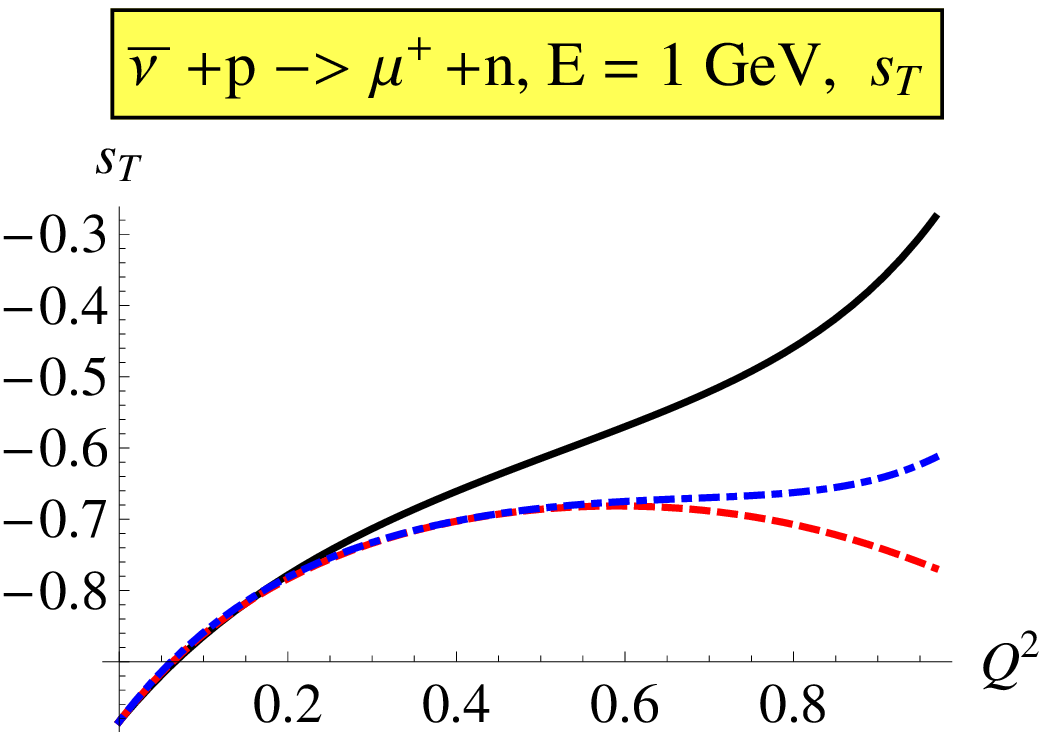}&
      \includegraphics[width=50mm]{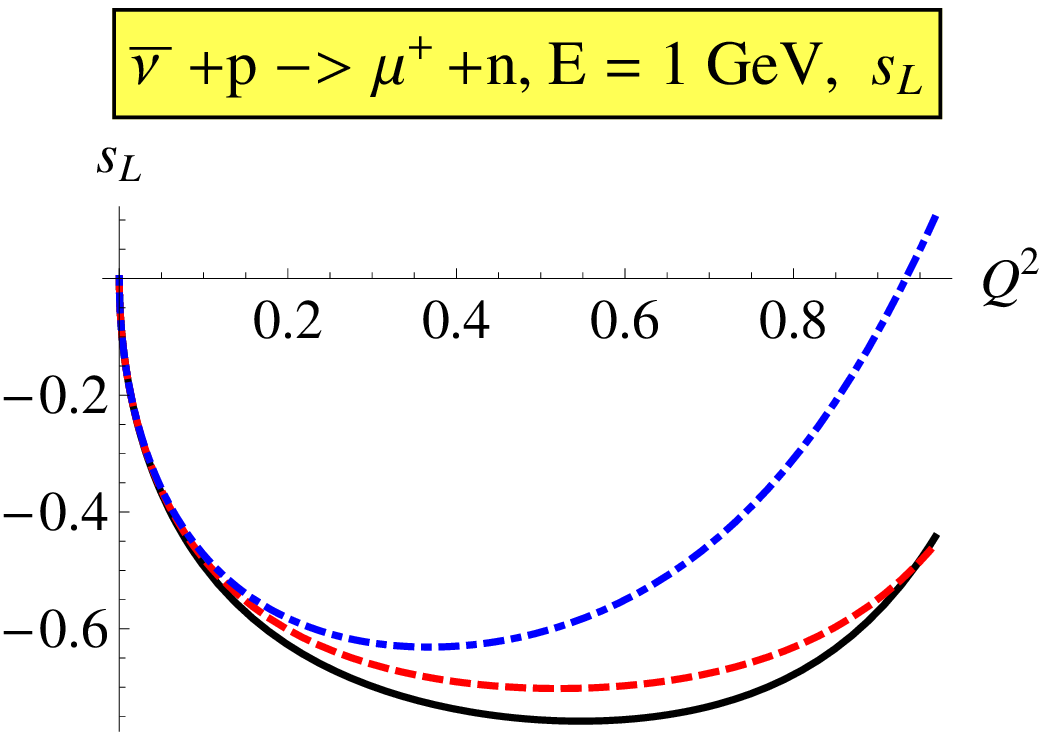}&
       \includegraphics[width=50mm]{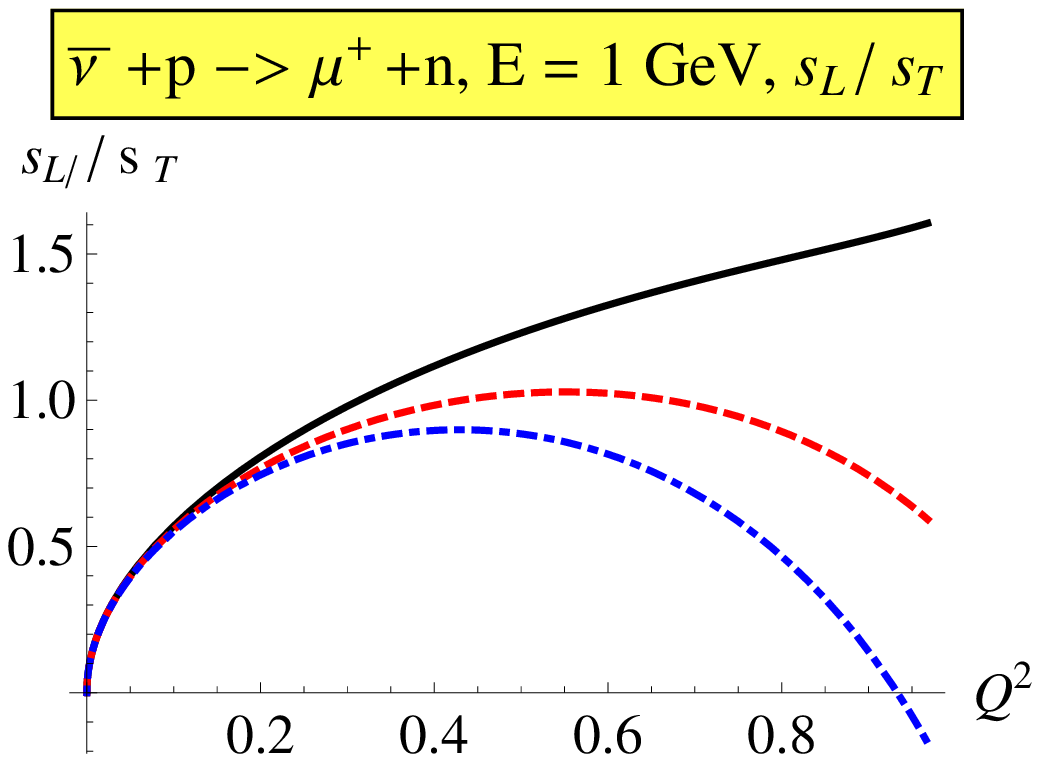} \\
    \textbf{(a)} & \textbf{(b)} & \textbf{(c)}
  \end{tabular}
  \caption{\footnotesize The
dependence of the transverse  $s_T$ and longitudinal $s_L$
polarizations  of the neutron at E=1 GeV (\textbf{(a)} and \textbf{(b)}), and their ratio $s_L/s_T$  \textbf{(c)}
on the values of $M_A$ (eq. (\ref{MA})) in $\bar\nu_\mu+p\to \mu^++n$.}
  \label{1}
\end{figure}

\begin{wrapfigure}[10]{R}{50mm}
  \centering 
  \vspace*{-8mm} 
  \includegraphics[width=50mm]{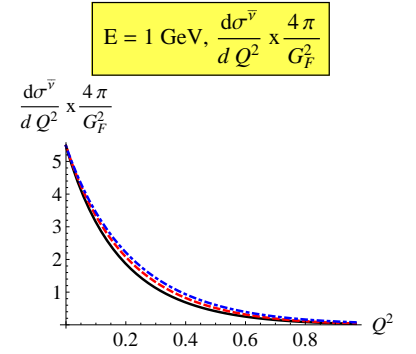}
  \caption{\footnotesize
The dependence on $M_A$ (see eq. (\ref{MA})) of the cross section of $\bar\nu_\mu+p\to
\mu^++n$  at E=1 GeV.}
  \label{2}
\end{wrapfigure}

Fig. (\ref{1}) shows that there is a clear sensitivity in the polarization of the final neutron in
$\bar \nu_{\mu}+p\to \mu^{+}+n$.
 It is most clearly pronounced for  $s_\|$ and, respectively, for the ratio
$s_\|/s_\perp$. An advantage of  $s_\|/s_\perp$ is that
many of the systematic uncertainties and radiative corrections cancel, however a magnetic field should be applied
to the detector in order to measure $s_\|$.
 This  sensitivity holds also for higher values of neutrino energies $E$. 
 In contrast,  Fig. (\ref{2}) shows that the cross section exhibits very weak sensitivity to $M_A$.

 There is almost no  sensitivity to the polarization of
 the proton in $\nu_{\mu}+n\to \mu^{-}+p$, but the polarizations are big and could present an independent measurement of $G_A$.

\end{document}